\documentclass[twocolumn,prb,floatfix,showpacs]{revtex4}

\usepackage{graphicx}

\newcommand{\crossprod}{\times}
\newcommand{\be}{\begin{equation}}
\newcommand{\ee}{\end{equation}}
\newcommand{\barr}{\begin{eqnarray}}
\newcommand{\earr}{\end{eqnarray}}
\newcommand{\breakeq}{\nonumber \\ &&}
\newcommand{\unit}[1]{\mathrm{\,\,#1}}
\newcommand{\diel}{\epsilon_0}

\begin{document}

\title{Light emission from Na/Cu(111) induced by a scanning
tunneling microscope}
\author{Peter Johansson}
\affiliation{Department of Natural Sciences,  University of \"Orebro,
          S--701\,82 \"Orebro, Sweden}
\author{Germar Hoffmann and Richard Berndt}
\affiliation{Institut f\"ur Experimentelle und Angewandte Physik,
          Christian-Albrechts-Universit\"at zu Kiel, D--24098 Kiel, Germany}

\date{\today}

\begin{abstract}
Measurements of light emission from a scanning tunneling microscope probing
a Na overlayer on the (111) surface of Cu are reported along with results
of a model calculation that essentially agree with the experimental ones.
The observed light emission spectra show two characteristic features
depending on the bias voltage.  When the bias voltage is smaller than the
energy of the second quantum well state formed outside the Na overlayer 
the light emission is due to a plasmon-mediated process,
while for larger biases light emission is mainly caused by quantum well
transitions between the two levels.
\end{abstract}


\maketitle

\section{Introduction}

The formation of quantum well states (QWS) near the Fermi level is
a characteristic feature of the electronic structure of alkali
overlayers on the (111) surfaces of noble metals.
\cite{Wallden_SSC80,lin:87,Diehl_SS96,Stampfl95,
Tochihara_PSS98,Hellsing_PRB00}  In these systems electrons may
be confined to a narrow surface region; they cannot escape from
the surface because of the vacuum barrier, and at the same time
they cannot propagate into the noble metal because the periodic
potential in the bulk creates a local band gap near the L point of
the Brillouin zone.  Alkali overlayers thus offer an interesting
opportunity to study confined electron systems in metals and they
have attracted considerable interest.

Up to now, most investigations of alkali overlayers have been
carried out using photoemission spectroscopy (PES), inverse
photoemission spectroscopy (IPES) and two-photon photoemission
(2PPE). \cite{lin:87,car:00,dud:91,fis:91,car:96} Thus the
energies of the quantum well states have been studied as a
function of substrate and overlayer material as well as overlayer
thickness (or coverage).
In addition to measuring the position in energy of the
quantum well states, the lifetimes of these states have also been
addressed.\cite{bauer:97,ogawa:99,borisov:01}
The quantum well states overlap in energy and
space with three-dimensional states, producing interesting
possibilities for quasiparticle decay to proceed simultaneously
through both two- and three-dimensional channels as well as
electron-phonon scattering.

In a recent experiment the system Na/Cu(111) was studied with an
alternative experimental technique, light emission induced by electron
injection from the tip of a scanning tunneling microscope (STM).
\cite{NaPRL} This technique has previously been used in studies of quite a
few systems, most notably noble metal surfaces and semiconductors.
\cite{review} In those cases photons are generated by a fraction of the
tunneling electrons that undergo inelastic tunneling processes in which an
amount of energy, limited to $eU$ with $U$ the bias voltage, is transferred
from the electron to the photon.  If the substrate is metallic, the rate of
spontaneous light emission is increased compared with the case of inverse
photoemission from an isolated surface.
\cite{gim:88,gim:89,joh:90,ber:91,per:92,Ueh:92,bergim:93,
joh:98,ueh:99,nil:00}
This is due to enhanced vacuum fluctuations of the electromagnetic field in
the tunnel gap between tip and sample as a result of the formation of
localized interface-plasmon modes there.  For semiconductor samples, on the
other hand, electromagnetic effects are less important, instead the light
emission is due to interband transitions in the semiconductor.
\cite{ALV90,CDS} The STM tip serves to locally inject or generate minority
charge carriers which then recombine with majority carriers giving rise to
luminescence. Similar studies have also been carried out on semiconductor
surfaces,\cite{thi:99,kag:01} semiconductor quantum wells\cite{ren:91} and
quantum dots,\cite{lind:96} and adsorbed molecules.\cite{ber:93,poi:01}

Interestingly enough, the light emission from Na/Cu(111) reported
in Ref.\ \onlinecite{NaPRL}, appears to involve both
mechanisms described above. For a wide range of coverages,  Na on
Cu(111) exhibits several QWS. The lowest one, QWS1, has an energy
close to the Fermi level (see Table \ref{Ei}). At the coverages to be
discussed below, it is either occupied or unoccupied. The other
state, QWS2, lies well above the Fermi energy. As long as the bias
voltage is low enough that electrons cannot be injected into QWS2
light emission is still possible, and it proceeds through the same
``plasmon-mediated'' mechanism as in the case of clean noble-metal
surfaces. In this case electrons undergo an inelastic transition
from a filled state in the tip to the empty QWS1 and a photon is
emitted.  This gives a fairly broad peak in the light emission
spectrum. Once the bias voltage is high enough that electrons can
be injected directly into the upper quantum well state (QWS2) the
emission mechanism changes.  A large part of the tunnel current
will now go through QWS2, and light emission will mainly be due to
transitions between QWS2 and QWS1, yielding a fairly sharply
peaked spectrum.  Of course, since the QWS wave functions are to a
large extent confined to the region of space between the tip and
sample, the light emission rate is still enhanced by the
electromagnetic fluctuations there.

In this paper we will support the scenario outlined above by model
calculations that lead to very good qualitative agreement between
experiment and theory. We will also present additional experimental
results.

The rest of the paper is organized in the following way.  In
Section
\ref{Expsec} we outline the experimental setup, and the experimental
results are presented in Sec.\ \ref{Measuresec}. Section
\ref{Theorysec} gives a comprehensive description of the model calculation
that we have carried out.
Next, in Sec.\ \ref{Discussec}, we compare
theoretical and experimental results and discuss the implications. The
paper is summarized in Sec.\ \ref{summarysec}.
Finally, an Appendix presents a derivation of the rate of light emission
based on the Keldysh non-equilibrium Green's function technique.

\section{Experimental details}
\label{Expsec}

The experiments were performed with an ultra-high vacuum (UHV) STM operated
at a temperature $T=4.6$~K\@.\cite{Diss_Kliewer} Photons
in the energy range $1.2 $ eV $ < $ h $\nu < 3.5$ eV were detected with a
lens-system in UHV coupling the light to a grating spectrometer and a
liquid nitrogen cooled charge-coupled device camera.\cite{GHaufbau:01}  All spectra have been
corrected for the wavelength dependent detection efficiency. Throughout the
measurements it was verified that the surface structure was not modified
during data acquisition.

W tips were prepared by electrochemical etching and subsequent
sputtering and annealing in UHV. The Cu(111) surface was cleaned by
repeated cycles of Ar-ion bombardment and annealing. Na films were
evaporated from outgassed SAES Getters sources onto the Cu
crystal held at room temperature. A
quartz crystal microbalance was used to estimate
the coverage $\Theta$ which was further calibrated by the known binding
energies of the quantum well states.
\cite{lin:87,car:00,dud:91,fis:91,car:96} After preparation at room
temperature the sample was transferred to the STM and cooled to
$T=4.6$~K.\@

Following Ref.~\onlinecite{tan:91} we define 1~monolayer (ML) as
the most densely packed structure of the first Na layer, namely a
$({3}/{2}\times {3}/{2})$ mesh. This pattern corresponds to 4 Na
atoms per 9 first layer Cu atoms.

\section{Experimental results}
\label{Measuresec}

\begin{figure}[htb]
  \includegraphics[angle=0, width=8 cm]{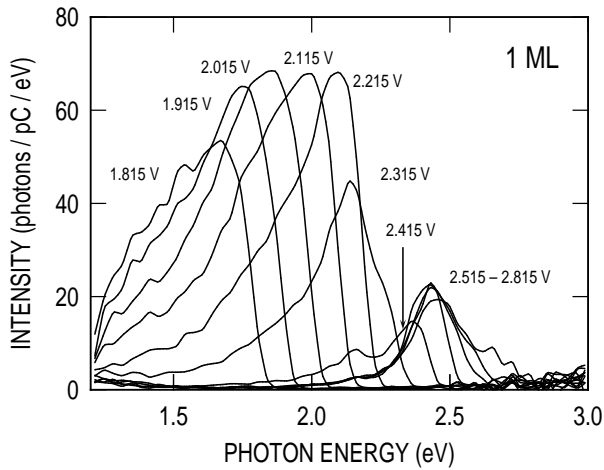}
\caption{
   Light emission spectra of 1 ML Na measured for a series of
   sample voltages
   between 1.815 V and 2.815 V with a tunneling current of 10 nA. Small
   apparent undulations at low ($ \lesssim 1.6$ eV) and high ($\gtrsim 2.6$ eV)
   photon energies are due to counting statistics. There is a qualitative
   change once the voltage reaches
   $\approx 2.5$ V: Below, the peak emission
   shifts with the
   sample voltage; above, the emission maximum remains at
   $h \nu \approx 2.4$ eV.}
\label{spexseries}
\end{figure}

Figure \ref{spexseries} shows fluorescence spectra recorded over a range of
sample voltages
$U$ from Cu(111) covered with 1 monolayer  of Na.  The
emission from these surfaces is comprised of two distinct components.
First, there is an emission band which is most clearly observed at low $U$.
Its maximum shifts linearly to higher energies as $U$  is increased. This
emission is similar to the plasmon-mediated emission observed from noble
metal surfaces due to inelastic tunneling processes.
\cite{gim:89,joh:90,ber:91,Ueh:92,ijn}

A new spectral structure,
the position of which only weakly depends on the
sample voltage $U$, emerges at h$\nu = 2.4$ eV
when the bias is raised to $U \gtrsim 2.5$ V.\@
From its intensity, assuming isotropic emission, a quantum efficiency of
roughly $10^{-5}$ photons per electron is estimated which is higher than
typical values for conventional inverse photoelectron spectroscopy.

\begin{figure}[htb]
  \includegraphics[angle=0, width=8 cm]{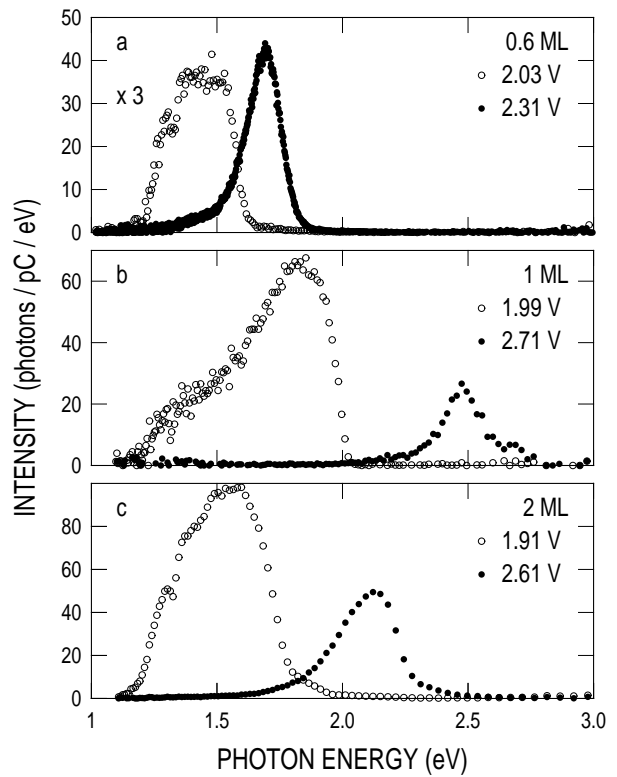}
\caption{
   Characteristic spectra, obtained with a low
   sample voltage (circles)
   for which injection of electrons into the upper quantum well state is
   not possible, and a high 
   sample voltage (dots) where electrons
   can be injected into the upper QWS.  The three panels display results
   from samples covered by 0.6, 1, and 2 monolayers of Na.  
   A small dip of the intensity at $\approx 1.3$ eV is due to a sharp
   absorption of the optical fiber used in the experiment, which is not
   fully corrected for.
   %a: 30 nA at 2.03 V, 180 nA at 2.31 V
   %b: 100 nA
   %c: 100 nA
}
\label{cover}
\end{figure}

Qualitatively similar observations were made at other coverages
(Fig.\ \ref{cover}).
In addition to a plasmon-related emission (circles) at low bias voltages
we find emission (dots) at higher photon energies which are
almost independent of the applied sample voltages.  The positions of these
emission features vary with the Na coverage:
$h \nu \approx$ 1.7, 2.5, and 2.1 eV
at coverages 0.6, 1, and 2 ML, respectively.
Comparing these photon energies to the
data of Table \ref{Ei} the emission is assigned to interband transitions
between quantum well states. Small deviations between the expected
transition energies ($E_2-E_1$) and the measured photon energies
are in part due to the electric field of the
tip which causes a Stark shift.\cite{becker,bin:85,Stark}
This shift is strongest for higher energy states.

Closer inspection reveals a distinct difference of the data from 0.6 and 2
ML compared to the 1 ML case.  At 1 ML coverage, a sharp drop on the high
energy side of the emission occurs owing to energy conservation given by
the condition $h \nu = e U$. The emission from the 0.6 and 2 ML Na films
exhibits a similarly sharp drop at a different energy, $h \nu = e U - E_1$,
where $E_1$ is the energy of quantum well state 1 relative to the Fermi
level.

\begin{table}[h]
\caption{\label{energy} Energies of quantum well states of Na
on Cu(111) in eV relative to the Fermi energy $E_F$ as obtained
from tunneling spectroscopy of the differential conductance.  At a coverage
of 1 monolayer, the lowest state is occupied ($E_1<0$).}
\begin{ruledtabular}
\begin{tabular}{r|rrr}
      & 0.6 ML & 1 ML  & 2 ML \\
\hline
$E_1$ &  0.4   & -0.15 & 0.15 \\
$E_2$ &  2.05  &  2.3  & 2.2  \\
\end{tabular}
\end{ruledtabular}
\label{Ei}
\end{table}

\section{Model and theory}
\label{Theorysec}

\subsection{General considerations}
\label{general:sub}

In this section we will use a rather simple, basically one-dimensional
model of the W-tip/Na/Cu(111) system that nevertheless captures the
essential physics of the experiment and yields calculated spectra that
can explain the experimental results.

The general framework of this calculation is
based upon what we have used
in earlier calculations\cite{joh:90,ZPB} of light emission from noble metal
surfaces.
As the more detailed derivation in the Appendix shows, the intensity of the
emitted light (per unit photon energy and solid angle) can be calculated
from the expression
\be
   \frac{dP}{d(\hbar\omega)d\Omega}
   =
   \frac{2\pi}{\hbar}
   \sum_{i,f}
   |j_{fi}|^2  \frac{|G(\omega)|^2}{(2\pi)^3}
   \frac{\hbar\omega^2}{2\epsilon_0 c^3}
   \delta(E_i-E_f-\hbar\omega).
\label{crossec1}
\ee
Thus, the intensity is found from a summation over filled
initial electron states $i$ and empty final states $f$,
and $j_{fi}$ is the current matrix element between these states,
\be
   j_{fi} =
   \frac{-ie\hbar}{2m}
   \int d^3r
   \left[
      \frac{\partial \psi_f^{*}}{\partial z} \psi_i
      -
      \psi_f^{*}
      \frac{\partial \psi_i}{\partial z}
   \right].
\label{jfi1}
\ee
In Eq.\ (\ref{crossec1}), $\diel$ is the permittivity of vacuum and $c$ the
speed of light, and $G$, finally, is an electromagnetic enhancement factor.
Note that Eq.\ (\ref{crossec1}) agrees with
Eqs.\ (6) and (7) of Ref.\ \onlinecite{ZPB}, considering that the formulas
in that work employed CGS units.

Since the calculation of $G(\omega)$ in Eq.\ (\ref{crossec1}) has been
treated in earlier papers (see, for example, Ref. \onlinecite{MCD2})
we will just give a brief outline of the procedure here.
As discussed in the Appendix, a reciprocity relation makes it possible to
interchange the source and detection points in the electromagnetic
calculation. Thus, since we are interested in evaluating the light emission
intensity found at an observation angle $\theta$=1 rad., we let an incident
electromagnetic wave hit the tip-sample system from exactly that direction
and calculate the electric field enhancement at a point between the tip
apex and the surface, where it reaches its highest values. The field in the
cavity between the tip and sample is essentially constant along the
direction normal to the surface, however, in the lateral direction the
field enhancement begins to drop off at distances exceeding 2--3 nm from
the symmetry axis.  The optical properties of tip and sample are modeled
using macroscopic dielectric functions for Cu and W found from Ref.\
\onlinecite{Palik}.  We have approximated the optical properties of the Na
layer by the Cu dielectric function since the treatment of such a thin
layer in terms of macroscopic dielectric functions would not be very
reliable.\cite{liebsch:01}  The tip is represented  by a sphere with radius
$R=30 \unit{nm}$.

\subsection{Model potential}

\begin{figure}[htb]
\includegraphics[angle=0, width=8 cm]{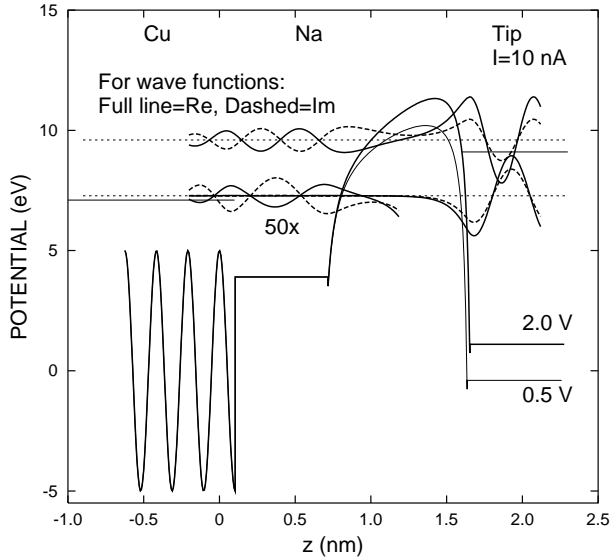}
\caption{
   Illustration of the model potential at two different values of the bias
   voltage. In addition the wave functions corresponding to the two quantum
   well states have been calculated using boundary conditions corresponding
   to an electron impinging on the barrier from the tip side when $U=2
   \unit{V}$.  Note that the wave function at the QWS2 energy penetrates
   the barrier with high probability, however, since this energy here lies
   above the tip Fermi energy, this does not contribute to the tunnel
   current.
}
\label{potfig1}
\end{figure}

We use a one-dimensional model potential, illustrated in Fig.\
\ref{potfig1}, to calculate both the tunneling
current and the matrix elements that set the light emission rate. The
parameter values that enter this model originate from the physical
properties of bulk Cu and bulk Na that are most relevant to the problem at
hand. Similar
models have been used in earlier calculations\cite{Smith,LW} (see
also Ref.\ \onlinecite{Zangwill}).

Of course, using a strictly one-dimensional model corresponds to a situation
where both electrodes, sample and tip, are completely flat. This is not
the real situation in an STM experiment. Thus, we
use the 1D model to calculate a current density which then multiplied by
a suitable effective area yields the tunnel current. A similar procedure is
applied for the matrix-element calculations as well.

Inside the copper sample the potential is modulated along the (111)
direction (normal to the surface) as
\be
   V_{\rm Cu} (z)
   =
   V_{111} (e^{igz} + e^{-igz})
   =
   2\, V_{111} \cos{(gz)},
\label{Cupot_eq}
\ee
where the reciprocal lattice wave vector $g=2\pi/a_{\rm Cu}$,
and $a_{\rm Cu}=2.08$
\r{A} is the inter-plane distance in Cu in the (111) direction. The
top layer is centered at $z=0$, and the copper sample is assumed to end at
$z=z_{\rm Cu}=a_{\rm Cu}/2$. The value of $V_{111}$ is taken to be 2.5
eV\@.\cite{LW}
The potential in (\ref{Cupot_eq}) yields a band gap for energies
(related to electron motion in the $z$ direction) between
$E_{g}-V_{111}$ and $E_{g}+V_{111}$, where $E_g=\hbar^2 (g/2)^2/(2m)
\approx 8.7$ eV is the kinetic energy of a free electron with a wave vector
at the L point on  the Brillouin zone boundary. Thus, just as in
copper our model has a 5 eV wide band gap at the L point. This feature is
crucial in forming quantum well states in the overlayer system.
Experimentally the energy difference between the lower band edge and the
Fermi level is 0.9 eV, we therefore  put the Cu Fermi level at
$E_F=E_g-V_{111}+0.9\unit{eV}=7.1\unit{eV}$.

We treat the sodium overlayer as a bulk, free-electron metal.
The thickness is assumed to correspond to 2 monolayers (6.13 \r{A}).\cite{LW}
The constant value of the potential in the sodium is set to
$V_{\rm Na}=3.9 \unit{eV}$. The choice of this value is based on the fact that
the Fermi energy of bulk Na is 3.2 eV, so that now an electron with the
kinetic energy of 3.2 eV in the Na layer will have a total energy of
$(3.9+3.2) \unit{eV}=7.1 \unit{eV}$, i.e.\ identical to $E_F$.

Moreover, in order to get a tunnel current
from electrons with energies in the Cu band gap, it is
necessary to add a negative imaginary part to the potential. We have
chosen to put this in the sodium
layer, thus the full Na potential is
$$V_{\rm Na} - i \Gamma, \ \ {\rm with}\ \  \Gamma=0.1 {\rm eV}. $$
$\Gamma$ in an approximate way represents the effects of electron-phonon,
electron-electron, and interface scattering that in the real system eventually
remove electrons from the quantum-well states near the surface.
If the potential had this value in all of space, the electron density
due to a particular state would decay in time as
$e^{-2\Gamma t/\hbar}$. Thus the lifetime would be
$\tau=\hbar/(2\Gamma)\approx 3 \unit{fs}$, and the peak in the spectral
function associated with the state would have a full width at half
maximum of $2\Gamma$.
It must be kept in mind that an electron in a QWS only spends part of
the time ($\approx 50$ \% for QWS1) in the Na layer.
Our choice for $\Gamma$ would thus give a lifetime of some 6 fs, and a line
width of 0.1 eV for QWS1.  These values are comparable with measured values
found in the literature.\cite{acarl:97}

We have for simplicity assumed that the potential is constant in the W tip.
This is of course a rather crude approximation, but not a
crucial one since the electronic structure of the tip is not of primary
importance for the analysis of the experiment at hand. With the
free-electron band width $W=8.0 \unit{eV}$, the
potential in the tungsten tip at zero bias is
$V_{\rm tip}=E_F-W=-0.9 \unit{eV}$, however, when the tip is biased the
potential is given by $V_{\rm tip} = E_F-W+eU$.

Finally, we need a barrier potential to use in vacuum between
the Na overlayer and the tip. This potential essentially consists of two
parts:  a tilted square barrier and image potential contributions.
The tilted square barrier can be written
\be
   V_{\rm tilt}(z) = \frac{z_{\rm tip}-z}{d} (E_F + \phi_{\rm Na})
      +
          \frac{z-z_{\rm Na}}{d} (E_F + eU + \phi_{\rm tip}),
\ee
where $z_{\rm tip}$ is the coordinate of the tip apex, $z_{\rm Na}$ is the
coordinate of the Na overlayer surface, and $d$ is the tip-sample
separation. Here  $\phi_{\rm Na}=2.7 \unit{eV}$  and $\phi_{\rm tip}=5.2
\unit{eV}$ denote the Na and tip work functions, respectively, and $U$ is
the bias voltage.  The image contributions are
\be
   V_{\rm im}(z) =
   - \frac{e^2}{4\pi\diel\,\, 4\, (z-z_{\rm Na}^{im})}
   - \frac{e^2}{4\pi\diel\,\, 4\, (z_{\rm tip}^{im}-z)}.
\ee
We have chosen the image plane position as
\be
   z_{\rm Na}^{im} -z_{\rm Na} =
   - {e^2} / [16\pi \diel (\phi_{\rm Na}+E_F-V_{\rm Na})]
\ee
in the sample. This guarantees that the total potential
$V_{\rm tilt}+V_{\rm im}$ equals $V_{\rm Na}$, the potential inside Na,
at $z=z_{\rm Na}$
provided the tip is far away.  In the tip, following the same reasoning,
\be
   z_{\rm tip}^{im} -z_{\rm tip} =
    {e^2} / [16\pi \diel (\phi_{\rm tip}+W)].
\ee
The resulting potential is illustrated in Fig. \ref{potfig1} for two
different values of the bias voltage.

The model potential described above yields an electronic structure of the
Na overlayer that is in reasonable agreement with experimental
observations. In particular, using this model we are able to reproduce the
essential features found in the experiment by Hoffmann, Kliewer, and
Berndt.\cite{NaPRL}  At the same time it should be pointed out that the
model is not detailed enough to reproduce exactly the same energy level
positions as found experimentally. Given that the Na layer is modeled using
bulk parameters, the theoretical results should be more accurate for thicker
overlayers.  We have therefore concentrated on calculating results for the
2-monolayer case.

\subsection{Wave functions}

To calculate the tunnel current and subsequently the light emission
intensity we must solve the Schr\"odinger equation for the electron wave
functions. Since we have chosen to  work with a potential that is
translationally invariant in the directions perpendicular to the tunneling
direction, we can write all wave functions on the form
\be
   \psi({\bf r}) = \psi(z) e^{i \vec{k}_{\|}\cdot\vec{r}_{\|}}.
\ee

In the copper sample, in view of the potential given by
Eq.\ (\ref{Cupot_eq}), we make the nearly-free electron model Ansatz
\be
   \psi(z) = \alpha e^{i k z} + \beta e^{i (k-g) z}
\ee
for the wave function. Inserting this into the Schr\"odinger
equation one finds that there is a band gap for energies (related to the
motion in the $z$ direction) between
$E_{g}-V_{111}$ and $E_{g}+V_{111}$.
In this energy interval there are no bulk
states in the copper, however, one can still have surface states
for which
$ k=p-iq$
with
$ p=g/2$,
and
$$
   q =
   \sqrt{2 m (\sqrt{4\varepsilon E_g + V_{111}^2} - E_g - \varepsilon)}
   / \hbar.
$$
Thus, the wave function envelope decays exponentially as $e^{qz}$ and the
electron is confined to the region near the copper surface.
Since we use a positive value for the corrugation of
the potential ($V_{111}=2.5$ eV) states near the bottom of
the gap will be p-like, whereas states near the top of the gap are
s-like.\cite{Smith}

The wave function in Na takes the form
\be
   \psi(z) = A e^{ik_{\rm Na} z} + B e^{-i k_{\rm Na} z},
\label{psiNa}
\ee
where
$ k_{\rm Na} = \sqrt{2 m (\varepsilon-V_{\rm Na}+i\Gamma)} / \hbar $
and the branch cut of the square root function is placed along the negative
real axis so that the imaginary part of $k_{\rm Na}$ is positive.
In the barrier region, the wave function must be integrated numerically and
then joined to the tip wave function
\be
   \psi(z) = C e^{i k_{\rm tip} z} + F e^{-i k_{\rm tip} z},
\label{psibarr}
\ee
(with $ k_{\rm tip} = \sqrt{2 m (\varepsilon-V_{\rm tip})} / \hbar $)
at $z=z_{\rm tip}$.
The coefficients $C$ and $F$, as well as $A$ and $B$ in Eq.\ (\ref{psiNa}),
must be determined by wave function matching.

\subsection{Tunnel current}

Before we can evaluate the matrix elements entering Eq.\
(\ref{crossec1}) we have to calculate the tunnel current $I_{\rm dc}$, or
rather, determine the tip-sample separation $d$ that yields a certain, set
value for $I_{\rm dc}$.
The probability current density in the $z$ direction associated with one
particular wave function can be written
\be
   j_{\psi} =
   \frac{1}{A_{\rm eff} L} \frac{1}{|F|^2}
   {\rm Re} [\psi^*(z) \frac{\hat{p}_z}{m} \psi(z)].
\ee
Here the first two factors serve as normalization. $A_{\rm eff}$ is the
effective area of the tunnel contact, which we also use as a normalization
area for the wave functions,\cite{Areanote} $L$ is a normalization length
(in the tip), and by dividing by $|F|^2$ we normalize the current to the
current carried by the wave impinging on the barrier from the tip side.
For a real-valued potential the probability current $j_{\psi}$ is
independent of the $z$ coordinate. In the present case, with a potential
that has a non-zero imaginary part in the sodium layer, the probability
current is independent of $z$ to the right (in the barrier and tip) of the
Na layer, and can be evaluated anywhere in that part of space.  $j_{\psi}$
vanishes, on the other hand, in the copper since we cannot have any
propagating states in the energy gap there.  A tunnel current is flowing
across the tunnel gap only because of the scattering processes that
eventually scatter electrons out of the quantum well states confined to the
Na overlayer and surface region of the copper.

The total electric current is obtained by summing over all contributing
states  to get a total current density and then multiply by the electron
charge and the effective tunneling area
\be
   I_{\rm dc}= -e A_{\rm eff} \sum_{\psi} j_{\psi}.
\label{Itot1}
\ee
This sum can then be turned into an energy integral.

\subsection{Intensity of emitted light}

To calculate the differential power of the emitted light we must carry
out the sum over initial and final electron states as indicated in Eq.
(\ref{crossec1}). In contrast to the case of light emission from noble
metals the final electron state is in this case, at least in one sense,
discrete. Light is  primarily emitted while the electron traverses the vacuum
barrier and then QWS1 is in practice the only possible final state
(assuming parallel momentum is conserved); it dominates the local density
of states completely in the energy range just above the substrate Fermi
level.

Therefore in the calculations, we have solved for a bound state to
represent $\psi_f$. This state is calculated with essentially the same
potential as was used for the tunnel current calculation, however, in this
case we set $\Gamma$ equal to zero, and determine an energy eigenvalue by
requiring that the wave function is decaying well inside the tunnel barrier.
In the $x$ and $y$ directions this bound state wave function is assumed to
behave like a plane-wave state with a certain momentum.  Furthermore, in
the following we will assume that this state is almost always unoccupied so
that tunneling electrons can make inelastic transitions into it and emit
light at the same time.  This last assumption is reasonable because even if
most of the tunnel current passes via QWS1, the lifetime of this state is
of the order 10$^{-14}$ s while with a tunnel current of 10 nA the delay
between each tunnel event is $1.6\times10^{-11}$ s.

In our one-dimensional model the momentum parallel to the interfaces and
the electron spin is conserved in the inelastic tunneling process.
The sum over initial and final states in Eq.\ (\ref{crossec1}) then
reduces to integrals over parallel and perpendicular momentum,
\begin{widetext}
\be
   \frac{dP}{d(\hbar\omega)d\Omega}
   =
   \frac{\omega^2 |G(\omega)|^2}{8\pi^2\diel c^3}
   \sum_{i,f}
   |j_{fi}|^2
   \delta(E_i-E_f-\hbar\omega)
   =
   2
   \frac{\omega^2 |G(\omega)|^2}{8\pi^2\diel c^3}
   A_{\rm eff} L
   \int \frac{dk_z}{2\pi}
   \int \frac{d^2k_{\|}}{(2\pi)^2}
   |j_{fi}|^2
   \delta(E_i-E_f-\hbar\omega).
\label{crossec2}
\ee
Here the integration over parallel momentum can be turned into an energy
integration over an interval starting at zero and ending at the maximum
energy $E_{\|,{\rm max}}$ that the electron can have due to the motion
parallel to the interfaces. This energy is given by the {\em tip} Fermi energy,
i.e.\ $E_F+eU$ minus the $z$ motion energy $E_f+\hbar \omega$ in the
initial state, i.e.\ $E_{\|,{\rm max}}=E_F+eU-E_f-\hbar\omega$.  We then get
\be
   \frac{dP}{d(\hbar\omega)d\Omega}
   =
   2
   \frac{\omega^2 |G(\omega)|^2}{8\pi^2\diel c^3}
   A_{\rm eff} L
   \frac{mE_{\|,{\rm max}} \Theta(E_{\|,{\rm max}})} {2\pi \hbar^2}
   \int \frac{dk_z}{2\pi}
   |j_{fi}|^2
   \delta(E_i-E_f-\hbar\omega),
\label{crossec3}
\ee
where $\Theta$ denotes a step function.  It remains to carry out the $k_z$
integration, and also this is, in view of the $\delta$ function in the
integrand, straightforward.  If we let $E_b$ stand for the band bottom
energy in the tip, one has ${\hbar^2 k_z^2}/{(2m)} = E_i - E_b$.  It is
then possible to show that
\be
   \delta(E_i-E_f-\hbar\omega)
   =
   \frac{
      \delta\left(k_z+\sqrt{2m(E_f+\hbar\omega-E_b)/\hbar^2}\right)
   }
   {
      \frac{\hbar}{m}
      \sqrt{2m(E_f+\hbar\omega-E_b)}
   },
\ee
and when this is inserted into Eq.\ (\ref{crossec3}) we get the final
result
\be
   \frac{dP}{d(\hbar\omega)d\Omega}
   =
   A_{\rm eff}
   \frac{e^2 \omega^2 |G(\omega)|^2}{64 \pi^4 \diel \hbar c^3}
   \frac{(E_F + eU - E_1 - \hbar \omega)
      \Theta(E_F + eU - E_1 - \hbar \omega)
   }{\sqrt{2m(E_1+\hbar\omega-E_b)}}
   \left|
      \int
      dz
      \left[
     \frac{\partial \psi_f^*}{\partial z} \psi_i
     -
     \psi_f^*
     \frac{\partial \psi_i}{\partial z}
      \right]
   \right|^2.
\label{finalintensity}
\ee
\end{widetext}
Note that we have made the replacement $E_f \rightarrow E_1$ in the last
equation because the final state here is identical to the lowest QWS.
The $z$ integration is limited to the vacuum part of space, since the
electromagnetic field enhancement is much higher there than in the tip and
sample.

\section{Results and Discussion}
\label{Discussec}

\begin{figure}[htb]
\includegraphics[angle=0, width=8 cm]{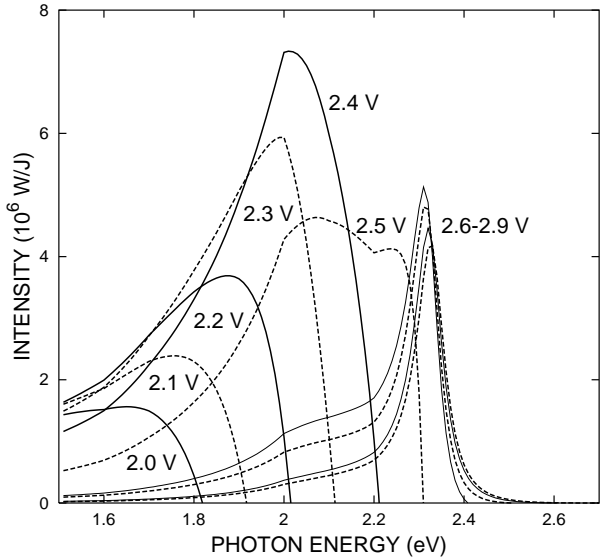}
\caption{
   Calculated light emission spectra for the Na/Cu(111) system with an
   overlayer thickness corresponding to 2 monolayers. The spectra were
   calculated for a series of bias voltages, while the tunnel current was
   kept fixed at 10 nA by varying the distance between tip and sample.
}
\label{calcfig}
\end{figure}

Figure \ref{calcfig} shows spectra calculated from the model of Sec.\
\ref{Theorysec} for a number of different bias voltages. The calculated
tunnel current was kept constant at $I_{\rm dc}=10 \unit{nA}$ in all cases
by changing the tip-sample separation.

We see that these spectra have a number of general qualitative features in
common with the experimental spectra shown in Figs.\ \ref{spexseries} and
\ref{cover}.  At first, for bias voltages $<2.5 \unit{V}$, the spectrum has
a relatively broad peak that is cut off on the high-energy side.  Since
QWS1 is the final state in the light emission process the maximum photon
energy  equals $eU-(E_{1}-E_F)$, and not $eU$, [cf.\ Eq.\
(\ref{finalintensity})].

Once the bias voltage becomes high enough that electrons can be injected
into the upper QWS most of the tunnel current will take this path.  At the
same time the tip-sample separation increases to maintain the tunnel
current at a constant value.  Quite naturally the electrons injected into
the upper QWS will also dominate the light-emission processes at these
higher voltages.  Most of the emitted photons will have an energy closely
corresponding to the energy difference between the two quantum well states.

Returning to the model, these facts can be understood as follows.
As long as the tip Fermi energy lies below QWS2 all initial state wave
functions will decrease exponentially upon traversing the tunnel barrier.
Therefore wave functions $\psi_i$, corresponding to a broad range of
energies give comparable contributions to the integral in
Eq.\ (\ref{finalintensity}). The shape of the spectrum in this case is
mainly determined by the phase space factor
$(E_F + eU - E_1 - \hbar \omega)$
and the field enhancement factor $|G(\omega)|^2$.

For higher bias voltages, when $E_F+eU$ exceeds $E_2$, it is instead the
last factor in Eq.\ (\ref{finalintensity}), the matrix element integral,
that determines the spectral shape.  The final state remains the
same and therefore, to have a large matrix element, the initial-state wave
function $\psi_i$ must be large in the part of space where the final state
resides. This will happen when the initial state energy coincides with
$E_2$ since then {\em resonant tunneling} into QWS2 is possible and
$\psi_i$ will look like the upper wave function illustrated in Fig.\
\ref{potfig1}.  Even if there are plenty of initial states with energies
both somewhat below and above $E_2$, they will not at all give as large
contributions to the emitted light intensity because their wave functions
are much smaller near the Na overlayer surface. As a result of this the
light emission spectrum develops a peak around
$h \nu = \hbar\omega=E_2-E_1$.

The overall intensity of the emitted light is smaller at the higher
voltages.  The main reason for this is that raising the voltage moves the
peak of the light emission spectrum away from the frequency range near 2 eV
where a tungsten tip and a copper sample have an interface-plasmon resonance
causing resonantly enhanced light emission. At a photon energy of 2.3--2.4
eV there is still a considerable field enhancement between the tip and
sample, but $G$ is typically down by a factor of 2 compared with the resonant
case.  Thus, the peak in the light emission spectrum at  
$h \nu \approx 2.3$
eV is entirely due to the special electron structure of the Na/Cu(111)
surface.

The experimental results also illustrate the last point in an interesting
way. With 1 ML coverage [Figs.\ \ref{spexseries} and \ref{cover}(b)] the
emission peak due to quantum well (or interband) transitions falls at a
high photon energy
where the electromagnetic enhancement is relatively small. Consequently,
comparing spectra taken at different voltages, the
plasmon-mediated light emission yields the more intense peaks in these
spectra.  However, for 0.6 ML coverage [Fig.\ \ref{cover}(a)] the quantum
well transition occurs at a lower photon energy, near the maximum of the
electromagnetic enhancement.  In this case the ``quantum-well peak'' is
more intense than the peak occurring at the lower bias voltage.

In this context let us discuss why it is reasonable to calculate the
electromagnetic enhancement at a point just below the tip apex both for
plasmon-mediated and QWS transition light emission. In the first case, this
is natural since the light emission event must take place while the
electron tunnels from tip to sample.  In the latter case, the electron is
injected into a fairly long-lived state, and in principle it may end up in
a point quite far away from the tip apex before a photon is emitted.
However, with a lifetime of 10 fs and a lateral electron velocity of $10^5$
m/s the distance traveled by the electron is no more than 1 nm, thus it
would still be in a region where the electromagnetic enhancement has not
dropped off very much
(cf.\ the discussion in subsection \ref{general:sub}).

From the calculated results, one can estimate the quantum
efficiency of the light emission process to be $10^{-5}$ emitted photons
per tunneling electron. This number compares well with the
experimental result. It is considerably larger than what is
observed for inverse photoemission processes at a single surface, however,
the quantum efficiency of STM-induced light emission from most notably Ag
samples may reach values between $10^{-4}$ and $10^{-3}$.

\section{Summary}
\label{summarysec}

In summary we have interpreted experimental observations of STM-induced
light emission from the quantum well system Na on Cu(111) using model
calculations of the electronic structure and the optical properties of the
tip-sample region.  The main features of the experimental data, namely two
distinct spectral structures, and their intensity variation with the tip-sample
voltage are reproduced by the calculations.  The emission lines are
attributed to (a) emission from a localized plasmon which is excited by
inelastic tunneling to a quantum well state and (b) transitions between two
quantum well states.  The electromagnetic enhancement present in the
tip-sample cavity substantially enhances the intensity of the emission and
explains the observed, high quantum efficiencies.

\begin{acknowledgments}
This research was supported by the European Commission via the TMR network
{\it EMIT}. The work of GH and RB  is further supported by the Deutsche
Forschungs\-gemein\-schaft via the ``Schwerpunktsprogramm
Elektronentransferprozesse an Grenzfl\"achen,'' and the work of PJ is
supported by the Swedish Natural Science Research Council (VR) and by the
SSF through the Nanometer Consortium at Lund University.
\end{acknowledgments}

% \appendix yields numbered (A, B, C,...) Appendixes.
% Use \appendix* to suppress numbering (just one Appendix)
\appendix*

\section{}
\label{appendix_sec}

In this appendix we derive Eq.\ (\ref{crossec1}) using
the Keldysh non-equilibrium Green's function  (GF)
technique.\cite{Mahan,Jauho}
We wish to calculate the intensity of the spontaneously emitted light at a
detection point ${\bf r}_0$ far away from the STM tip and sample.  Since
all photons appearing at this point that are of interest to us originate
from the region around the STM tip, the intensity can be found by
multiplying the electromagnetic energy density, excluding the zero-point
energy, by the speed of light. The energy density can be expressed in terms
of Keldysh GF's as
\be
   {\mathrm w} =
   i\diel
   \int_0^{\infty} \frac{d\omega}{2 \pi}
   [
      D_{EE}^{<}({\bf r}_0,{\bf r}_0,+\omega)
      +
      D_{EE}^{>}({\bf r}_0,{\bf r}_0,-\omega)
   ],
\label{Poynting}
\ee
where
$D_{EE}^{<}({\bf r}_0,{\bf r}_0, \omega)$
and
$D_{EE}^{>}({\bf r}_0,{\bf r}_0, \omega)$
are the Fourier transforms of
\be
   D_{EE}^{<}({\bf r}_0,{\bf r}_0, t)
   =
   - i
   \langle
   E_{\theta}({\bf r}_0, 0)
   E_{\theta}({\bf r}_0, t)
   \rangle
\label{Deedef1}
\ee
and
\be
   D_{EE}^{>}({\bf r}_0,{\bf r}_0, t)
   =
   - i
   \langle
   E_{\theta}({\bf r}_0, t)
   E_{\theta}({\bf r}_0, 0)
   \rangle,
\label{Deedef2}
\ee
respectively. In writing these expressions we have assumed that the tunnel
current causing light emission flows in the $z$ direction near the origin
and sends out p-polarized light with an electric field
${\bf E}=\hat{\theta}E_{\theta}$, pointing in the $\theta$ direction in the
far field.  Consequently, the radiated differential power can be written
\be
   \frac{dP}{d(\hbar\omega)d\Omega}
   =
   \frac{ic\diel r_0^2}{2 \pi \hbar}
   [
      D_{EE}^{<}({\bf r}_0,{\bf r}_0,+\omega)
      +
      D_{EE}^{>}({\bf r}_0,{\bf r}_0,-\omega)
   ].
\label{crossec4}
\ee
The two Keldysh GF's  in Eq.\ (\ref{crossec4}) actually yield identical
contributions and in the following we will only deal with $D^<$.

We need to find an expression for the contributions to
$D_{EE}^{<}$ that result from interactions between the electromagnetic
field and the electron system in the STM tip and sample. This interaction
is to lowest order
\be
   H' =
   \frac{e}{2m}
   \sum_n
   [
      {\bf A}({\bf r}_n)\cdot{\bf p}_n
      +
      {\bf p}_n\cdot{\bf A}({\bf r}_n)
   ],
\ee
where ${\bf A}$ is the electromagnetic vector potential
(${\bf E} = -\partial{\bf A}/\partial t$), the sum runs over the electrons,
and ${\bf r}_n$ and ${\bf p}_n$ are the electron coordinates and momentum
operators, respectively.  By performing an S-matrix expansion of
$D_{EE}^{<}$ to second order in $H'$ we find, using the rules for
``analytic continuation'',\cite{Jauho} that $D_{EE}^{<}$ can be expressed
as a product of two photon Green's functions and a current-current Green's
function,
\barr
   &&
   D_{EE}^{<}({\bf r}_0,{\bf r}_0, \omega)
   =
   \frac{1}{\hbar^2}
   \int d^3r_1
   \int d^3r_2
   \breakeq
   \times
   D_{EA}^{r}({\bf r}_0, {\bf r}_1, \omega)
   \Pi^{<}({\bf r}_1, {\bf r}_2, \omega)
   D_{AE}^{a}({\bf r}_2, {\bf r}_0, \omega).
   \breakeq
\label{DeeSmat}
\earr
These Green's functions are the Fourier transforms of
a retarded photon GF
\be
   D_{EA}^{r}({\bf r}, {\bf r'}, t)
   =
   -i \theta(t)
   \langle
   [
      E_{\theta}({\bf r}, t)
      ,
      A_{z}({\bf r'}, 0)
   ]
   \rangle,
\ee
an advanced photon GF
\be
   D_{AE}^{a}({\bf r}, {\bf r'}, t)
   =
   i \theta(-t)
   \langle
   [
      A_{z}({\bf r}, t)
      ,
      E_{\theta}({\bf r'}, 0)
   ]
   \rangle,
\ee
and the current-current GF
\be
   \Pi^{<}({\bf r}, {\bf r'}, t)
   =
   - i
   \langle
   j_z({\bf r'},0)
   j_z({\bf r},t)
   \rangle.
\ee
The current density operator
\be
   j_z({\bf r})
   =
   \frac{-ie\hbar}{2m}
   \left[
      \Psi^{\dag}({\bf r})
      \frac{\partial \Psi({\bf r})}{\partial z}
      -
      \Psi({\bf r})
      \frac{\partial \Psi^{\dag}({\bf r})}{\partial z}
   \right],
\ee
where $\Psi$ and $\Psi^{\dag}$ are electron annihilation and creation
operators.\cite{Rick}

At this point we can make a number of approximations and simplifications.
The inelastic tunneling events occur in a very small part of space in the
tunnel gap in a region of (sub-)nanometer size.  The photon Green's functions
do not vary very much on this length scale, so $D^r$ and $D^a$ can be taken
outside the integral, and ${\bf r}_1$ and ${\bf r}_2$ can be replaced by a
fixed point ${\bf r}_s$ in the tunnel gap in these functions.
Fourier transformation turns time derivatives into frequency
multiplications. Therefore the photon Green's functions only need to
involve the vector potential. Moreover, in Fourier space retarded and
advanced GF's are each others complex conjugates, and finally the
reciprocity theorem of electrodynamics allows us to interchange the source
and field points in the retarded photon GF. This gives us
\be
   D_{EA}^{r}({\bf r}_0, {\bf r}_s, \omega)
   D_{AE}^{a}({\bf r}_s, {\bf r}_0, \omega)
   =
   \omega^2
   | D_{z\theta}^{r}({\bf r}_s, {\bf r}_0, \omega) |^2.
\label{DDprod}
\ee

In a case where parts of space is filled with materials characterized by a
relative dielectric function $\epsilon_r({\bf r})$
the (tensor) photon Green's function $D_{\alpha\beta}$
solves\cite{Mahan,StatPhys2}
\be
   \left[
      \nabla \crossprod
      \nabla \crossprod
      \,
      -
      \,
      \epsilon_r({\bf r}) \frac{\omega^2}{c^2}
   \right]
   D_{\alpha\beta}^r ({\bf r}, {\bf r'}, \omega)
   =
   - \hbar \mu_0 \hat{\beta} \delta^3({\bf r}-{\bf r'})
\ee
and yields the $\alpha$ component of the vector potential in the point
${\bf r}$ if there is a $\delta$ function current source pointing in the
$\beta$ direction in ${\bf r'}$.
Thus,
$D_{\alpha\beta}^r ({\bf r}, {\bf r'}, \omega)$ can be calculated within
the framework of classical electrodynamics. The $z\theta$ element
of interest here can be written
\be
   D_{z\theta}^r ({\bf r}_s, {\bf r}_0, \omega)
   =
   -
   \frac{\hbar}{\diel c^2}
   \frac{e^{ikr_0}} {4\pi r_0}
   G(\theta_0,\omega),
\label{Dret}
\ee
where $G$ is an enhancement amplitude. $G$ is given by
the $z$ component  of the local electric field in the tunnel gap when a
plane wave of unit amplitude incident from the direction of ${\bf r}_0$
impinges on the tip-sample system. In free space one would simply have
$G(\theta, \omega)=\sin{\theta}$.  The detailed scheme for calculating $G$
has been described earlier, for example in Ref.\ \onlinecite{MCD2}.

It remains to evaluate the current-current Green's function
$\Pi^{<}({\bf r}_1, {\bf r}_2, \omega)$
and to carry out the integrations over the coordinates ${\bf r}_1$ and
${\bf r}_2$ in Eq.\ (\ref{DeeSmat}).
A straightforward evaluation yields the result
\barr
   &&
   \int d^3r_1
   \int d^3r_2
   \Pi^{<}({\bf r}_1, {\bf r}_2, \omega)
   =
   \breakeq
   =
   -2 \pi i \hbar
   \sum_{fi} |j_{fi}|^2
   \delta(\hbar \omega + E_f-E_i)
\label{Piint}
\earr
where the sum runs over filled initial electron states $i$ with energy
$E_i$ and wave function $\psi_i$  and empty final
electron states $f$ with energy $E_f$ and wave function $\psi_f$, and
\be
   j_{fi} =
   \frac{-ie\hbar}{2m}
   \int d^3r
   \left[
      \frac{\partial \psi_f^{*}}{\partial z} \psi_i
      -
      \psi_f^{*}
      \frac{\partial \psi_i}{\partial z}
   \right].
\label{jfi2}
\ee

By inserting the results of Eqs.\ (\ref{DeeSmat}), (\ref{DDprod}),
(\ref{Dret}), and (\ref{Piint}) in Eq.\ (\ref{crossec4}) we arrive at the
final result
\be
   \frac{dP}{d(\hbar\omega)d\Omega}
   =
   \frac{\omega^2\, |G(\theta_0, \omega)|^2}{8\pi^2 \diel c^3}
   \sum_{fi} |j_{fi}|^2 \delta(\hbar\omega+E_f-E_i),
\label{crossec5}
\ee
which is identical to Eq.\ (\ref{crossec1}).


\begin{thebibliography}{0}
\bibitem{Areanote} We used an effective area of 8 nm$^{2}$ in the
calculations. The light emission results are not very sensitive to the
exact value of $A_{\rm eff}$ as long as the tunnel current is kept at a
fixed value.
\end{thebibliography}
\end{document}